\documentclass[letterpaper,english,twocolumn,aps,prl,groupedaddress]{revtex4-1}
\usepackage[T1]{fontenc}
\usepackage[latin9]{inputenc}
\usepackage{units}
\usepackage{amsmath}
\usepackage{amssymb}
\usepackage{graphicx}
\usepackage{wasysym}
\usepackage{braket}

\def\kr{k_{\rm R}}                            				% kr
\def\Er{E_{\rm R}}                            				% kr
\def\th{t_{\rm h}}
\def\Wh{\Omega_{\rm h}}
\def\Wc{\Omega_{\rm c}}

\makeatletter

\pdfpageheight\paperheight
\pdfpagewidth\paperwidth

\makeatother

\usepackage{babel}
\begin{document}

\title{Spatial coherence of spin-orbit-coupled Bose gases}

\author{Andika Putra, F. Salces-C\'{a}rcoba, Yuchen Yue, Seiji Sugawa, I.
B. Spielman}

\affiliation{Joint Quantum Institute, National Institute of Standards and Technology,
and University of Maryland, Gaithersburg, Maryland, 20899, USA}

\date{\today}
\begin{abstract}
Spin-orbit-coupled Bose-Einstein condensates (SOBECs) exhibit two new phases of matter, now known as the stripe and plane-wave phases. 
When two interacting spin components of a SOBEC spatially overlap, density modulations with periodicity given by the spin-orbit coupling strength appear. 
In equilibrium, these components fully overlap in the miscible stripe phase, and overlap only in a domain wall in the immiscible plane-wave phase.
Here we probe the density modulation present in any overlapping region with optical Bragg scattering, and observe the sudden drop of Bragg scattering as the overlapping region shrinks.
Using an atomic analogue of the Talbot effect, we demonstrate the existence of long-range coherence between the different spin components in the stripe phase and surprisingly even in the phase-separated plane-wave phase.
\end{abstract}
\maketitle
% !TEX root = StripePhase.tex
% Informs TeXShop to look one folder up for the main file. 

Systems with coexisting order parameters, such as ferromagnetic superconductors~\cite{MatthiasPRL1958FerrSup}, supersolids~\cite{Pomeau1994prl}, or topological Kondo insulators~\cite{Dzero2010prlKondo}, exhibit rich phases with novel phenomena.
Spin-orbit coupled Bose-Einstein condensates (SOBECs) have a complex phase diagram including both ``stripe'' and ``plane-wave'' phases.
The stripe phase is expected to have coexisting order parameters~\cite{Stanescu2007prb,Lin2011nat,Ho2011} with supersolid-like properties~\cite{Li2013} marked by long-range phase coherence and periodic density modulations (confirmed by optical Bragg scattering~\cite{Li2017nat}) simultaneously present.
In contrast, the ``plane-wave'' phase behaves like a ferromagnetic spinor Bose-Einstein condensate (BEC), where its true many-body ground state is predicted to be massively entangled with application to precision magnetometry~\cite{Higbie2004,Stanescu2008}.  
In both the stripe and plane-wave phases, we readout a matter wave Talbot interferometer with optical Bragg scattering to detect coexisting periodic density modulations (long range diagonal order) and system-wide phase coherence (long-range off-diagonal order).
Unexpectedly, both phases exhibit both types of order.

\begin{figure}
\includegraphics{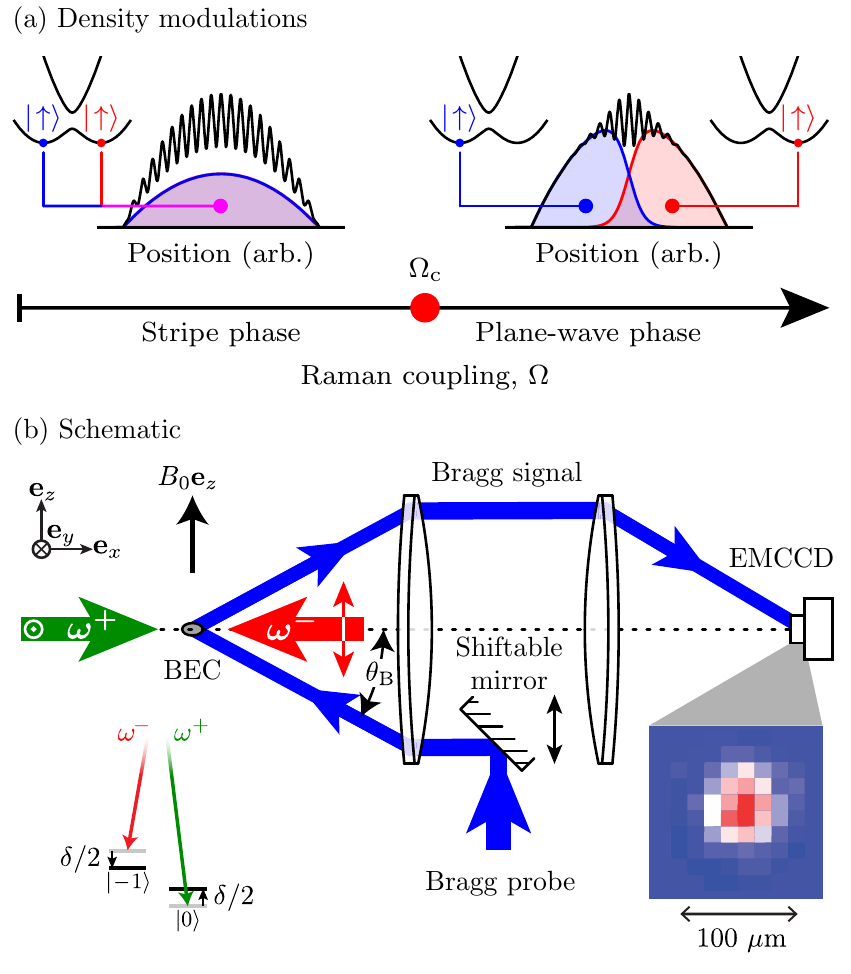}
\caption{Experimental concept and setup. (a) Schematic description of small-$\delta$ phase diagram with equal spin populations showing the stripe and plane-wave phases.  The spatial distribution of the two spin-orbit coupled spin states are marked in red and blue for $\ket{\uparrow}$ and $\ket{\downarrow}$ respectively, while the total density is in black.
The insets depict the dispersion of these states. 
(b) Laser configuration for realizing SOC system with two-photon Raman transition and detecting scattered Bragg signal from the stripe phase.  
We choose a bias magnetic field $B_0 \approx 20\ {\rm G}$.
The inset shows an example of diffracted Bragg signal as imaged by an EMCCD camera.}
\label{fig:Setup_stripe}
\end{figure}

Figure~\ref{fig:Setup_stripe}a schematically depicts the stripe and plane-wave phases of SOBECs, showing two salient features~\cite{Wang2010prl,Ho2011,Lin2011nat}: (1) system-wide periodic density modulations are associated with fully coexisting spin components in the stripe phase; and (2) highly localized density modulations are present at a domain-wall delineating phase-separated spin components in the plane-wave phase. Initial experiments with Raman coupled $^{87}$Rb Bose-Einstein condensates (BECs) identified these phases in terms of the degree of spatial overlap of the two spin components~\cite{Lin2011nat}, but not the microscopic density modulations.
Direct observation of these modulations in $^{87}$Rb BECs is challenging both because the $\approx\unit[400]{nm}$ modulation period is below the resolution of even the best quantum gas microscope~\cite{Bakr2009nat}, and the modulation contrast is small. Here we probe these modulations in long-lived equilibrium systems in both the stripe and plane-wave phases.

Our manuscript is organized as follows: (1) we introduce the physics of SOBECs; (2) we describe our experimental setup; (3) we cross-check our Bragg measurements with established techniques; (4) we demonstrate the coexistence of diagonal and off-diagonal order in the same system; and (5) we discuss the implications of these measurements on the issues of supersolidity in stripe-phase SOBECs.
 
{\it SOBECS with Raman coupling} We realized SOBECs described by the single-particle Hamiltonian
\begin{equation}
\hat{H}_{0}=\frac{\hbar^{2}}{2m}\left[\left(q_{x}-\kr\hat{\sigma}_{z}\right)^{2}+k_{\perp}^{2}\right]+\frac{\delta}{2}\hat{\sigma}_{z}+\frac{\Omega}{2}\hat{\sigma}_{x},\label{eq:SOCsingle}
\end{equation}
for particles of mass $m$.  
Here, $\delta$ and $\Omega$ describe Zeeman shifts from longitudinal and transverse fields respectively; and the spin-orbit coupling (SOC) strength $\kr$ defines the relevant energy scale $\Er=\hbar^{2}\kr^{2}/2m$.  
$\hbar q_{x}$ is the quasi-momentum along ${\bf e}_{x}$; $\hbar k_{\perp}$ is the linear momentum in the ${\bf e}_{y}-{\bf e}_{z}$ plane; and $\hat{\sigma}_{x,y,z}$ are Pauli operators.
The insets to Fig.~\ref{fig:Setup_stripe}a show the characteristic double-well dispersion associated with SOC, with minima separated by approximately $2\kr$, and energy gap equal to $\Omega$.
In our experiments we use two-photon Raman transitions to introduce the SOC term: the Raman laser wavelength determines the SOC strength $\kr=2\pi/\lambda_{\rm R}$; the Raman laser intensities determine $\Omega$; and the laser frequency differences imbue detuning $\delta$ to the SOC system~\cite{Lin2011nat,Ho2011}.

We describe the two spin-components of our system by the spinor wavefunction
$\left(\psi_{\uparrow},\;\psi_{\downarrow}\right)^{T}$, where the mean-field interaction energy density is
\begin{align*}
\varepsilon\!&=\!\left[\frac{c_0}{2}\!+\! \frac{c_2}{4}\right]\!\left[\left|\psi_{\uparrow}\right|^{2}\!\!+\!\left|\psi_{\downarrow}\right|^{2}\right]^2\!\!\!\!-\!\frac{c_2}{4}\!\left[\left|\psi_{\uparrow}\right|^{4}\!\!-\!\left|\psi_{\downarrow}\right|^{4}\right]
\!+\!\frac{c_2}{2}\!\left|\psi_{\uparrow}\psi_{\downarrow}\right|^{2}\!\!.
\end{align*}
Here $c_0$ and $c_2$ describe the inter- and intra-spin interaction parameters respectively, and $\bar{n}$ is the mean density.   
For dilute Bose-gases (with chemical potential $\mu\ll\Er$), the impact of interactions can be parameterized in terms of a scaled recoil energy $\Er^\prime = \Er + \mu / 4$; in this case the spin mixed, stable ground-state stripe phase, exists in a very narrow range of parameters~\cite{Lin2011nat}: with $\delta$ between $0$ and $c_2 \bar{n}/2$; and $|\Omega| < \Wc$, with the critical coupling strength $\Wc = 4 \Er^\prime\sqrt{-2 c_2/c_0}$.   As depicted in Fig.~\ref{fig:Setup_stripe}a (left), the stripe-phase density 
\begin{align*}
\frac{n(x)}{\bar{n}}=1+\frac{\Omega}{4\Er^\prime}\cos\left[k(\Omega)x+\phi\right],
\end{align*}
is modulated with wave-vector
\begin{equation}
\frac{k(\Omega)}{2 \kr}=\left[1-\left(\frac{\Omega}{4\Er^\prime}\right)^{2}\right]^{1/2}.\label{eq:periodicity}
\end{equation}
The phase $\phi$ describing the stripe's location~\cite{Ho2011,Li2012prl} results from the pre-existing phase difference between the two spin components along with the relative phase between the Raman laser beams. On the contrary, for the plane-wave phase ($|\Omega|> \Wc$) shown in Fig.~\ref{fig:Setup_stripe}a (right), density modulations are expected only within the domain-wall separating the now polarized spin components.

{\it Experimental setup} We produced $N\!=\!2.2(3)\!\times\!10^{5}$ condensed $^{87}\mathrm{Rb}$ atoms in a harmonic trap with frequencies $\left(f_{x},f_{y},f_{z}\right)=\unit[\left(105,67,40\right)]{Hz}$ and chemical potential $\mu\!=\!\unit[h\times1.46(20)]{kHz}$. 
Two Raman lasers, counter-propagating along ${\bf e}_{x}$, coupled the $\left|\downarrow\right\rangle\!\equiv\!\left|f=1,m_{F}=-1\right\rangle $ and $\left|\uparrow\right\rangle\!\equiv\!\left|f=1,m_{F}=0\right\rangle $ hyperfine levels of $^{87}\mathrm{Rb}$ $5{\rm S}_{1/2}$ electronic ground states.  
We used the tune-out wavelength~\cite{Arora2011pra} $\lambda_{\rm R}=\unit[790.034(7)]{nm}$ for our Raman lasers which defined the single-photon recoil energy $\Er=h\times\unit[3.678]{kHz}$.
 
% According to Andika's thesis the magnification of the imaging system is 3.9.  And the pixel size is
% 13 um, giving a magnified size of 3.3 um.  The thesis states that the initial ROI was 68 pixels wide
% but I count 17 pixels, but 4*17 = 68, so there is 4x binning.  So each displayed pixel is 13.3 um in size
% and there are 10 such pixels.  So that gives 133 um.  100/13.3 = 7.5.  So I will count 7.5 pixels and make 
% a 100 um scale bar.

We used optical Bragg scattering~\cite{Weidemuller1995,Birkl1995prl,Miyake2011} to detect periodic density modulations.
The Bragg probe laser, with wavelength $\lambda_{\rm B}=\unit[780.24]{nm}$, was $\approx\unit[6.3]{GHz}$ red-detuned from the $f=1\rightarrow f'=0,1,2$ transition within the $\mathrm{D}_{2}$ line~\cite{Muller2005pra}.
This put the Bragg probe beam in the far-detuned limit with respect to: the $\approx\unit[6]{MHz}$ transition linewidth, the $\approx\unit[10]{MHz}$ Zeeman shifts, and the $\approx\unit[300]{MHz}$ excited state hyperfine structure.  
In this limit the atomic susceptibility is almost entirely real and state-independent.
Figure~\ref{fig:Setup_stripe}(a) shows our experimental setup, with atoms located at the focus of a Keplerian imaging system aligned along ${\bf e}_{x}$.
The Raman lasers propagated along ${\bf e}_{x}$ and the Bragg probe had an incident angle $\theta_{\rm B}$ with respect to the optical axes.
A shiftable mirror in the back focal plane tuned $\theta_{\rm B}$ from $\unit[80]{mrad}$ to $\unit[280]{mrad}$, allowing the detection of Bragg scattering from structures with period from about $\unit[391]{nm}$ to $\unit[405]{nm}$; we used  $\theta_{\rm B}\approx\unit[0.2]{rad}$ in these experiments \footnote{
Due to the modest transverse size of our BEC, the Bragg peak was both broadened and shifted to an increased incoming angle~\cite{Slama2005pra}. 
We observed the lowest order Bragg peak at $\theta_{\rm B}\simeq\unit[0.2]{rad}$, larger than the theoretically predicted Bragg angle $\theta_{\rm B} =\cos^{-1}\left(\lambda_{B}/d\right)=\unit[0.16]{rad}$ for an infinite crystal with periodicity $d = \lambda_{\rm R}/2$. This shift is consistent with our numerical simulations.}.
In $^{87}{\rm Rb}$, the interaction constants~\cite{Widera2006} are $(c_0,c_2) = (779, -3.61)\times\unit[10^{-14}]{Hz\ cm^{3}}$, so the stable ground-state stripe phase was present for $\Omega\apprle0.21\Er$ and $\unit[-3.3]{Hz}<\delta/h<\unit[0]{Hz}$.

The Bragg diffracted signal, as shown in the inset of Fig.~\ref{fig:Setup_stripe}(b), was detected with an electron-multiplying charge-coupled device (EMCCD) camera.  As described in the supplementary material (SM), we first calibrated our Bragg signal using an optical lattice and found that the signal-to-noise ratio (SNR) of one occurred for a fractional density modulation of $\eta = 0.06$, providing practical detection threshold.

We prepared our SOBECs from an initial BEC with equal superposition of spin $\left|\uparrow\right\rangle$ and $\left|\downarrow\right\rangle$ at a desired detuning $\delta$, and linearly increased $\Omega$ from $0$ to $\Wh$ in $\unit[50]{ms}$.  
We then allowed the system to equilibrate for a hold time $\th$. 
At the transition from stripe to plane-wave phase at $\Wc\approx0.21\Er$, the expected density modulation contrast is just $\eta=0.045$: just below our detection threshold. 
Inspired by Ref.~\onlinecite{Hart2015nat}, we rapidly ramped $\Omega$ to $\approx7\Er$ in $\unit[200]{\mu s}$ just prior to our Bragg measurement, increasing $\eta$ to $\approx0.85$ (see SM).
This rapid ramp was slow compared to the $\approx4\Er$ energy spacing between the two branches of the SOC dispersion, but fast compared to the much slower many-body dynamics.
As a result, this process simply magnified the amplitude of the SOC driven stripes wherever they were present in the system.
We then turned the Raman lasers off and pulsed the Bragg laser with duration $t_{\rm B}$ ranging from $\unit[20]{\mu s}$ to $\unit[100]{\mu s}$.

\begin{figure*}
\includegraphics{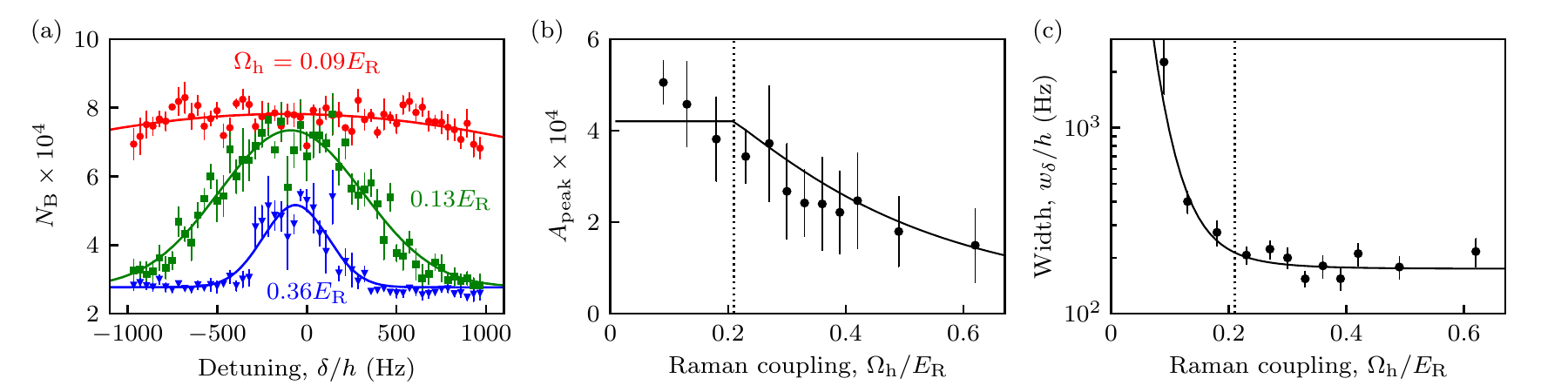}
\caption{
Bragg-scattering as a function of $\Wh$ and $\delta$.
(a) $N_{\rm B}\left(\delta\right)$ from a $t_{\rm B} = \unit[100]{\mu s}$ pulse for various coupling strengths $\Wh$; each data point is an average of more than 4 realizations.
The solid curves depict Gaussian fits to the data from which the peak amplitude $A_{{\rm peak}}$ and width $w_\delta$ in (b) and (c) are derived.
The increased background level as compared to Fig.~\ref{fig:Setup_stripe} is from an increased overall atom number.
In (b) and (c) the vertical dotted line mark the critical coupling strength at $\Wc=0.21\Er$, showing that the Bragg amplitude lacks a sharp feature at $\Wc$, while the width drops rapidly leading up to $\Wc$. 
% Background = 2.7662e4
}
\label{fig:Ome_Det_phDiag}
\end{figure*}

{\it Validation of method} We began by demonstrating our ability to maintain balanced spin mixtures very near $\delta=0$, in the process cross-checking our Bragg measurements against earlier TOF (time-of-flight) experiments~\cite{Lin2011nat}.
We characterized the transition from the stripe to plane-wave phase as a function of Raman coupling $\Wh$ and detuning $\delta$.
Figure~\ref{fig:Ome_Det_phDiag}(a) shows the number of photo-electrons $N_{\rm B}$ in our detection region as a function of $\delta$ at different values of $\Wh$ for a fixed hold time $\th=\unit[1]{s}$.
We observe Bragg scattering in a narrow detuning window that decreases in width and amplitude as $\Wh$ increases. 

Figure~\ref{fig:Ome_Det_phDiag}(b) quantifies the amplitude in terms of the peak height $A_{{\rm peak}}$ obtained from Gaussian fits to $N_{\rm B}\left(\delta\right)$.
We might expect the Bragg scattering amplitude to be constant in the stripe phase $\Wh < \Wc$ where the spin components mix, and then to vanish in the plane-wave phase when the gas becomes locally polarized.  
However, even when different plane-wave regions phase separate, density modulations are present in the domain wall separating the different phases, allowing some Bragg scattering. 
The spin healing length $\xi_{\rm s} / \xi = \left(\Wh^2/\Wc^2-1\right)^{-1/2} c_0/(-c_2)$ in terms of the conventional healing length $\xi = \hbar/\sqrt{2 m \mu}$.
$\xi_{\rm s}$ sets the domain wall size~\cite{Lin2011nat} and diverges at $\Wc$.  
Figure~\ref{fig:Ome_Det_phDiag}(b) shows $A_{{\rm peak}}\left(\Wh\right)$ rapidly falling with increasing coupling strength, consistent with the expected trend. 
The solid curve is a fit to our scattering model model (derived from the above reasoning and developed in the SM) with the overall Bragg signal as the only free parameter.
This model shows only qualitative agreement with data, a point we will return to shortly.

Figure~\ref{fig:Ome_Det_phDiag}(c) plots the Gaussian width $w_\delta$.
Even for $\Wh<\Wc$, a small detuning $\delta\neq0$ that breaks the degeneracy of the two spin states can cause the initially spatially mixed states to relax into a polarized gas in the lower energy spin state: a plane-wave phase with no Bragg scattering.  
When $\Wh=0$ there are no spin-changing processes, and the spatially mixed state is stable indefinitely, independent of $\delta$.
The width is thus large for small $\Wh$ (slower spin relaxation) and decreases as $\Wh$ increases (faster spin relaxation).
The width has no marked feature at $\Wc$, and is well fit by a power-law~\cite{Lin2011nat}, here $a(\Wh/E_{E})^{-4} + w_{\infty}$. 
This indicates that the process by which the spin population polarizes in the presence of detuning is dependent on the Raman coupling strength, but not the initial zero-detuning phase.

In all cases, the detuning window is far wider than the $\unit[3.3]{Hz}$ range of detuning where the stripe phase is thermodynamically stable.  This is as expected: the timescale for the spin populations to reach the expected equilibrium population can be in excess of several seconds for small detunings (see Ref.~\onlinecite{Lin2011nat} and SM for a discussion of the equilibration timescale).  In what follows we focus on near-zero detunings that lie within this meta-stable region and where the physics is governed by $\Wh$ alone.  

\begin{figure}
\begin{centering}
\includegraphics{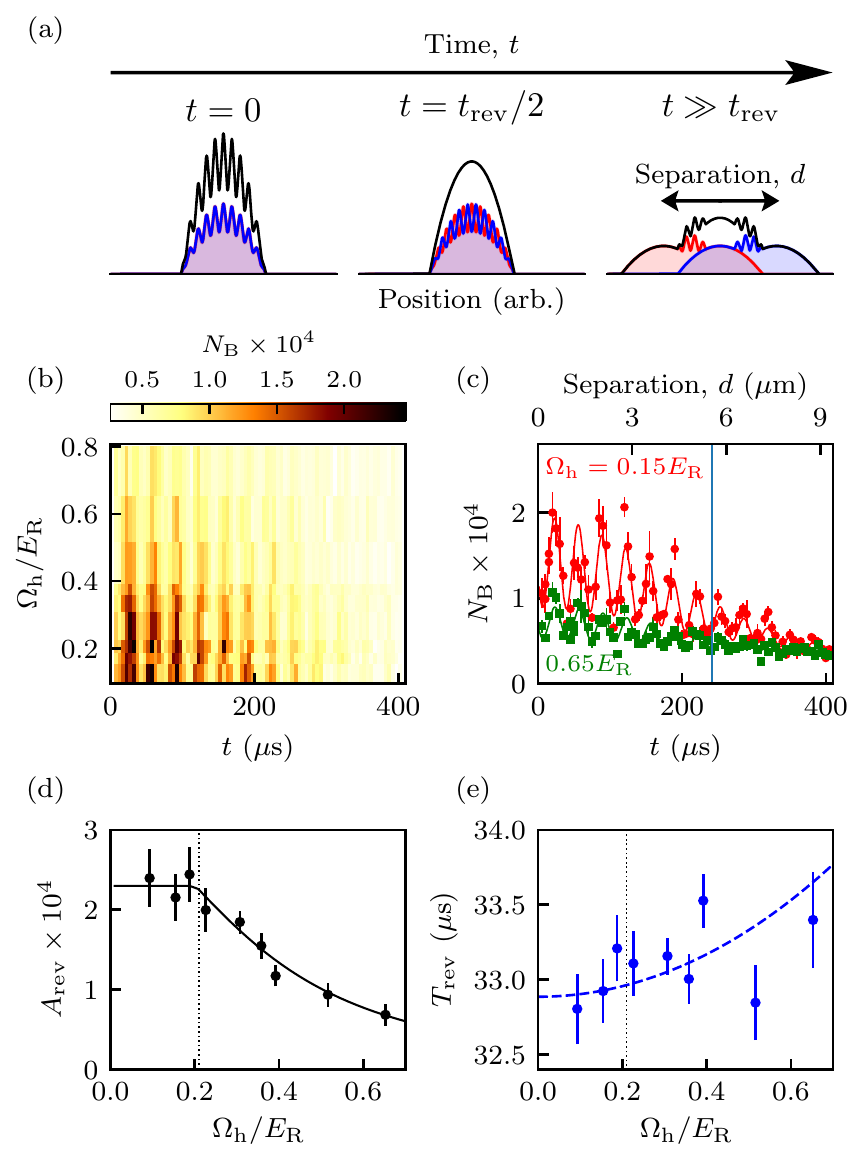}
\end{centering}
\caption{
Periodic revival of Bragg signal at $\delta=\unit[0]{Hz}$.
(a) Schematic representation of evolution of stripes during free evolution.
The black, red, and blue curves depicts total density, spin-up density and spin-down density respectively.
(b) Observed Bragg counts $N_{\rm B}$ from a $t_{\rm B} = \unit[20]{\mu s}$ pulse as functions of $\left(\Wh,t\right)$.
(c) $N_{\rm B}\left(t\right)$ for various coupling strength $\Wh$ showing revivals characteristic of an atomic Talbot effect.
The solid curves are joint fits of the model described in the text with shared parameters: decay time $t_{\rm d} = \unit[250]{\mu s}$ and background level $c=3487$ counts.
The vertical blue line depicts the separation equal to the calculated Thomas-Fermi radius.
(d) and (e) depict the amplitude $A_{{\rm rev}}$ and period $T_{{\rm rev}}$ obtained from fits to the data in (b).  The vertical dotted lines show the predicted transition strength at $\Wc$.  The dashed blue curve indicates prediction for $T_{{\rm rev}}(\Omega)$ shifted down by $\unit[1.1]{\mu s}$.
} 
\label{fig:Ome_Trev_phDiag}
\end{figure}

{\it Spatial coherence} Finally, we present our main observation demonstrating the spatial coherence of the SOBECs. 
Here we altered our measurement procedure to include a free evolution time $t_{{\rm rev}}$ following the turn off of the Raman lasers but prior to the Bragg pulse.  During this time, the different spin and momentum components that comprised the Raman dressed states underwent free evolution creating a matter-wave Talbot interferometer~\cite{Talbot1836paper,Miyake2011,Santra2017ncomm}.
A coherent matter-wave with wave-vector $\kr$ exhibits a coherence revival after a time period of $T_{{\rm rev}}=h/8\Er=\unit[34.0]{\mu s}$, during which time momentum components traveling with velocity $\pm 2 \hbar \kr / m$ separated by a distance $\lambda_{\rm R}$.
Figure~\ref{fig:Ome_Trev_phDiag}(a) schematically depicts this behavior: the left panel shows modulations in total density (black) and in each spin component (red and blue) at $t=0$; the center panel shows that after $T_{{\rm rev}}/2$ the modulation pattern in each spin component moved $\pm 1/4$ of the overall modulation period, yielding a flat density profile.
The right panel shows the long-time behavior in which the spin components moved a distance comparable to the overall system size.

The periodic revivals in Fig.~\ref{fig:Ome_Trev_phDiag}(b) occured very near the $\unit[34]{\mu s}$ free-particle Talbot time, only about one-third of our earlier $\unit[100]{\mu s}$ Bragg pulse time.
This indicates that all of our previous measurements inadvertently integrated over about three periods of collapse and revival.
To resolve the Talbot signal, we largely mitigated this effect by reducing the pulse time to $t_{\rm B} = \unit[20]{\mu s}$, and averaging over at least four experimental realizations to account for the reduced signal present in each measurement.

Figure~\ref{fig:Ome_Trev_phDiag}(c) shows $N_{\rm B}$ as a function of $t_{{\rm rev}}$ for a range of $\Wh$, each constituting a single horizontal cut through Fig.~\ref{fig:Ome_Trev_phDiag}(b). 
In Fig.~\ref{fig:Ome_Trev_phDiag}(c), we observe damped oscillatory behavior that provides a lower bound to the coherence length of the system (other physical effects~\cite{Miyake2011} may also cause the decay of $N_{\rm B}\left(t_{{\rm rev}}\right)$). 
Our observations are complicated by the $\unit[20]{\mu s}$ Bragg pulse which is not short compared to the revival time.  
We modeled the integrated Bragg signal as a sinusiod with Gaussian decay~\footnote{We also considered exponential decay, but the overall $\chi^2$ was increased by a factor of four.} convolved with our Bragg pulse to obtain
\begin{align*}
N_{\rm B}\left(t\right) & = A_{{\rm rev}}\int_{t}^{t+t_{\rm B}}\frac{{\rm d}t'}{t_{\rm B}}\cos^{2}\left(\frac{\pi t'}{T_{{\rm rev}}}\right)e^{-(t'/t_{\rm d})^2}  + c,\label{eq:fittingNrevival}
\end{align*}
as displayed by the solid curves in Fig.~\ref{fig:Ome_Trev_phDiag}(c).
Here $t_{\rm B}=\unit[20]{\mu s}$ is the Bragg pulse duration and the fitting parameters are revival amplitude $A_{{\rm rev}}$, revival period $T_{{\rm rev}}$, decay time $t_{\rm d}$, and constant $c$.

Figure~\ref{fig:Ome_Trev_phDiag}(d)-(e) shows the revival amplitude $A_{{\rm rev}}$, and period $T_{{\rm rev}}$, as a function of coupling strength $\Wh$. 
The amplitude $A_{{\rm rev}}$ gradually decreases above $\Wh>0.21\Er$,
which we attribute to the onset of phase separation and subsequent
increasing separation between the two plane-wave components.
The solid curve depicts the fit to the scattering model described in the SM with the overall scattering strength as the only free parameter, showing near perfect agreement with experiment.
Allowing $\Wc$ to vary in the scattering model produces a value $\Wc=0.20(1)$, also in agreement with our expectations.
Figure~\ref{fig:Ome_Trev_phDiag}(e) shows revival periods close to $T_{{\rm rev}}=\unit[33]{\mu s}$, just below the naive single-particle prediction.
Our model in Eq.~\eqref{eq:periodicity} predicts an increase in $T_{{\rm rev}}$ for larger $\Wh$ as the stripe wave-vector $k(\Omega)$ falls.
This increasing trend is plotted by the blue dashed curve; both this model and the null hypothesis are consistent with the data.

Lastly, the decay time $t_{\rm d} = \unit[250]{\mu s}$ was independent of $\Wh$, indicating that the transition from the stripe phase to the plane-wave phase was not associated with any decrease in spatial coherence.  During this $\unit[250]{\mu s}$, the interfering momentum are states separated by $\unit[5.8]{\mu m}$, comparable to the $R_{\rm TF} = \unit[5.5]{\mu m}$ Thomas-Fermi radius [shown by the vertical line in Fig.~\ref{fig:Ome_Trev_phDiag}(c)].  We conclude that the system was fully coherent even in the phase-separated plane-wave phase.

{\it Implications for supersolidity} As has now been observed with dipolar atoms~\cite{Natale2019}, a traditional supersolid is a phase of matter with two broken symmetries~\cite{Boninsegni2012}: the broken gauge symmetry of a BEC (giving a superfluid phonon mode) and the broken translation symmetry of a lattice (giving a separate lattice-phonon mode).  
On one hand, we confirmed that diagonal order is present~\cite{Li2017nat}, and demonstrated that this coexists with off-diagonal order: a supersolid?  On the other hand, a BEC in a shallow optical lattice has off-diagonal order, with density modulations (diagonal order) simply imprinted by the lattice potential~\cite{Greiner2002}: not a supersolid.

With the Raman lasers off, our system is a two-component spinor BEC with two broken symmetries giving an overall phase (giving a superfluid phonon mode), and a relative phase between the spin components (giving a spin-wave mode); translational symmetry is unbroken: not a supersolid.
Adding Raman coupling continuously connects this spinor phase to the stripe-phase.
The modulation period (from Eq.~\ref{eq:periodicity}) is externally imposed by the Raman lasers, with spatial phase set both by the relative phase between the Raman lasers and the pre-existing relative phase between spin components.  Similar to the lattice case, no new symmetries are broken and no new collective modes are created: not a supersolid?  
Although no new symmetries are broken, the spin-wave mode acquires an inertial contribution from the periodic density modulations inducing a gap at the edge of the associated Brillouin zone: as would be expected of a super-solid's lattice-phonon mode~\cite{Li2013}.
We conclude that this system some properties with conventional supersolids, but is best given its own name: the super-stripe phase, as suggested in Ref.~\onlinecite{Li2013}. 
The lattice-phonon mode remains undetected, and its observation would be a true smoking gun for observation of super-stripes.

\begin{acknowledgments}
This work was partially supported by the AROs atomtronics MURI, the AFOSRs Quantum Matter MURI, NIST, and the NSF through the PFC at the JQI. 
We are grateful for the very thoughtful and detailed eleventh hour reading of our manuscript by Qiyu Liang and Alina Pinero.
\end{acknowledgments}

\bibliographystyle{apsrev4-1}
\bibliography{StripePhase}

\end{document}

\title{Supplemental Materials: Spatial coherence of spin-orbit-coupled Bose gases}

\author{Andika Putra, F. Salces-C\'{a}rcoba, Yuchen Yue, Seiji Sugawa, I.
B. Spielman}

\affiliation{Joint Quantum Institute, National Institute of Standards and Technology,
and University of Maryland, Gaithersburg, Maryland, 20899, USA}

\date{\today}
\maketitle

%%%%%%%%%% Merge with supplemental materials %%%%%%%%%%
% \widetext

\makeatletter
\renewcommand{\theequation}{S\arabic{equation}}
\renewcommand{\thefigure}{S\arabic{figure}}
\renewcommand{\bibnumfmt}[1]{[S#1]}
\renewcommand{\citenumfont}[1]{S#1}
%%%%%%%%%% Prefix a "S" to all equations, figures, tables and reset the counter %%%%%%%%%%

This Supplemental Materials addendum contains three sections.  Firstly we describe our technique for calibrating the Bragg signal using an optical lattice potential.  Secondly, we compute the expected contrast of density modulations as a function of Raman coupling strength $\Omega$ and wave-vector $q$.  Lastly, we describe our 1D model of system-wide Bragg scattering including the signal from a domain wall delineating different spin components.

\section{Bragg calibration}

We first calibrated the Bragg signal using a 1D spin-dependent lattice~\cite{McKay2010njp}. We adiabatically loaded spin-$\left|\downarrow\right\rangle $ condensates into the lattice potential, giving a density modulation along ${\bf e}_{x}$ with periodicity $d=\lambda_{\rm R}/2\approx \unit[395]{nm}$. 
As the lattice depth $V_{0}$ increases, the periodic modulation of the atomic wavefunction grows and eventually separates into well resolved Wannier functions associated with each lattice site~\cite{Marzari2012rmp}.
However, in the weak lattice limit, we obtain the photon counts corresponding to the Bragg diffraction from a simple sinusoidal density modulation.

\begin{figure}
\includegraphics{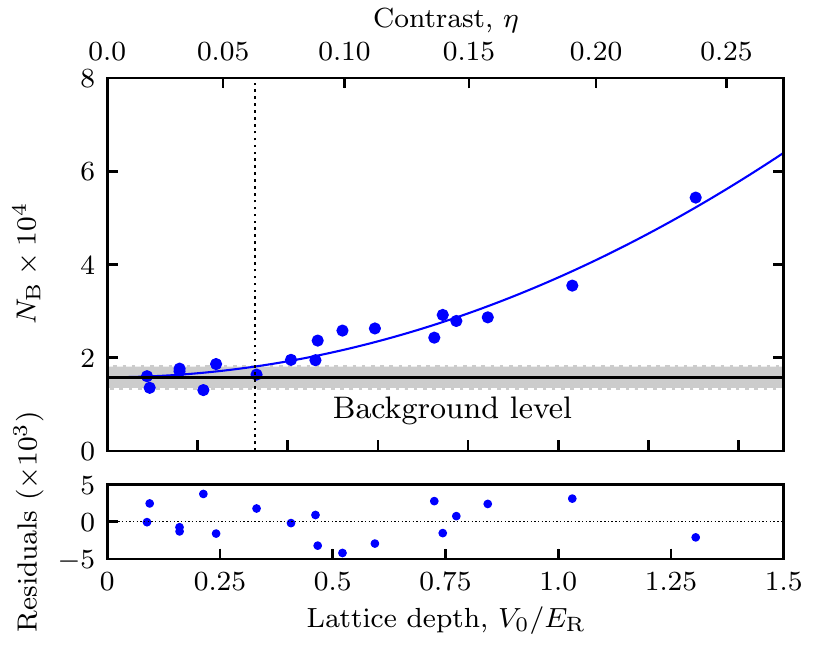}
\caption{Lattice calibration data. Photon counts $N_{\rm B}$ from a $t_{\rm B} = \unit[100]{\mu s}$ Bragg pulse as a function of lattice potential depth $V_{0}$ and density contrast $\eta$. Solid curve is a quadratic fit to the data giving the $N_{\rm B}=(2.1\times10^4)\eta^2 + 1.6\times10^4$, where $1.6\times10^4$ is the background level from off-resonant scattering even without any lattice.
The bottom plot shows the resulting residuals, with standard deviation of $\approx 2\times10^3$ counts.}
\label{fig:sm:lattice}
\end{figure}

Figure~\ref{fig:sm:lattice} shows the total photoelectron counts $N_{\rm B}$ as a function of $V_{0}$ at low lattice depth. 
We determine our lattice depth $V_{0}$ with Kapitza-Dirac diffraction~\cite{Gadway2009oe} and calculate its corresponding fractional density contrast $\eta$ by numerically solving the Gross-Pitaevskii equation~\cite{Bao2003scaling}. 
Fitting to the expected quadratic behavior allows us to identify the ultimate sensitivity of our method and quantifies the $N_{\text{BG}}=1.6\times10^{4}$ background counts present even without coherent Bragg scattering.
The fit residuals give a standard deviation of $2.3\times10^{3}$ giving a signal-to-noise ratio (SNR) of one at $\eta = 0.06$ (corresponding to $1.13$ times the background level), where the lattice depth is $0.3 \Er$.  
We therefore identify $\eta = 0.06$ as a practical detection limit.

\section{Modulation depth}

Here we compute the contrast of density modulations when two different eigenstates $\ket{\pm q, -}$ of the spin-orbit coupled Hamiltonian matrix
\begin{align}
H(q) &= \left(\begin{array}{cc} q^2 & 0 \\ 0 & q^2\end{array}\right) + \left(\begin{array}{cc}-2 q & \Omega/2 \\\Omega/2 & 2 q\end{array}\right)
\end{align}
are simultaneously present.  Here $q$ is the quasi-momentum quantum number and $-$ denotes the lower branch of the spin-orbit coupled dispersion, and we are working in dimensions of the recoil energy $\Er$ and recoil momentum $\kr$.  We first explicitly compute $\ket{\psi} = (\ket{+q, -} + \ket{-q, -})/\sqrt{2}$ as a vector in the basis of the spin states $\ket{\uparrow,\downarrow}$ and then evaluate the associated spatial density to identify any modulations.

The explicit eigenvectors of the matrix $H(q)$ are
\begin{align*}
\psi_{{\bf q},-}(x) =& \sqrt{\frac{1}{(\Omega/2)^2 + \left(q + \sqrt{q^2 + (\Omega/2)^2}\right)^2}}\times\\
&\left(\begin{array}{c}\left[q + \sqrt{q^2 + (\Omega/2)^2}\right] e^{i(q+1)x}  \\ -\left[\Omega/2\right]e^{i(q-1)x}\end{array}\right)\\
\equiv& \sqrt{\frac{1}{b_q^2 + a_q^2}}\left(\begin{array}{c}a_q e^{i(q+1)x}  \\ - b_q e^{i(q-1)x}\end{array}\right).
\end{align*}
The superposition of $\psi_{+q,-}(x)$ and $\psi_{-q,-}(x)$ gives the total probability as a function of position
\begin{align}
p(x) &= 1 + \left[1 + \left(\frac{4q}{\Omega}\right)^2\right]^{-1/2} \cos(2 q x).
\end{align}
This reduces to the expression for modulation amplitude in the main body when $q$ is taken to be that of the ground-state of the SOC dispersion.  

\begin{figure}
\begin{centering}
\includegraphics{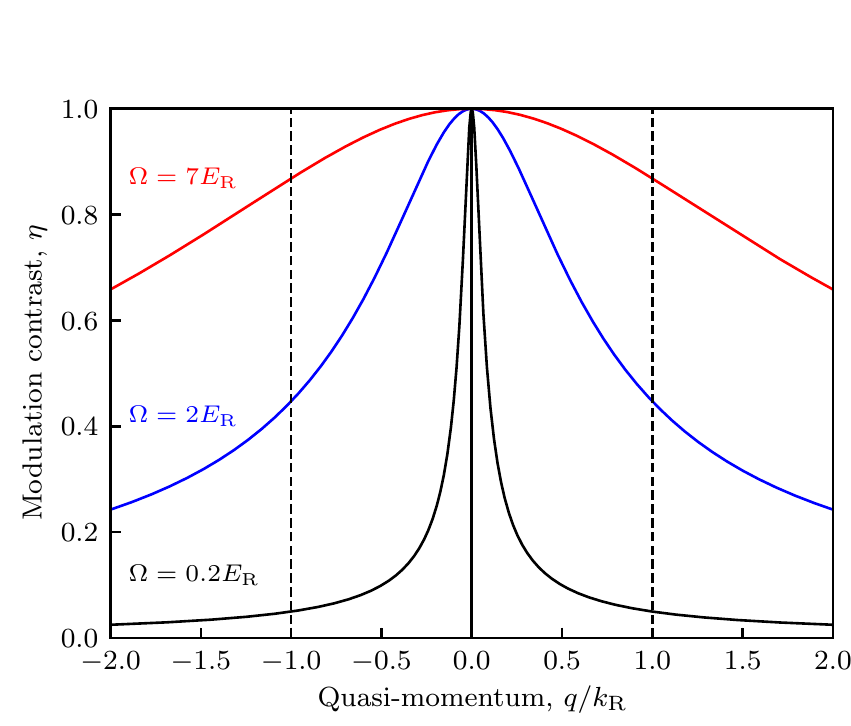}
\end{centering}
\caption{
Density modulation contrast plotted as a function of $q$ for three different final coupling strengths (black: $\Omega = 0.2\Er$, blue: $\Omega = 2\Er$, and red $\Omega = 7\Er$).  At $\Omega=7\Er$ this shows the near-uniform and nearly full contrast present between the $q/\kr =\pm 1$ relevant for our experiment, marked by the vertical dashed lines.
} 
\label{fig:sm:modulation}
\end{figure}

In our experiment, we rapidly ramped up $\Omega$ to about $7\Er$ just prior to our measurement.  This process is instantaneous compared to the large-scale spatial structure governing the overall distribution of spins in the system, but adiabatic compared to the local SOC energy scales.  As a result, at any place where stripes are present, the local contrast is given by the final $\Omega$.  This is analogous to band-mapping methods used in optical lattice experiments.  Figure~\ref{fig:sm:modulation} plots the expected modulation strength for a range of $\Omega$, and shows that for the range of quasi-momentum present in our experiment $q/\kr =\pm 1$ (near the local minima of the SOC dispersion), the contrast is always above $0.8$ for $\Omega=7\Er$.

\section{Equilibration time}

\begin{figure}[t]
\begin{centering}
\includegraphics{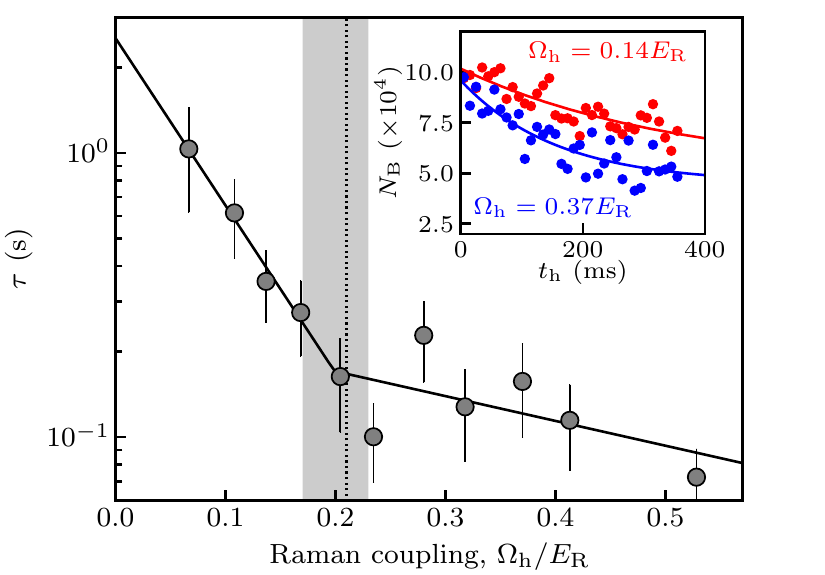}
\end{centering}
\caption{
Decay of Bragg signal at $\delta=\unit[0]{Hz}$. 
Inset: $N_{\rm B}$ from a $t_{\rm B} = \unit[100]{\mu s}$ pulse as a function of $\th$ for $\Wh<\Wc$ (red) and for $\Wh>\Wc$ (blue). 
Solid curves depict exponential fits to the data. 
Main panel: Lifetime $\tau$ of the Bragg signal as a function of $\Wh$.  
The solid curve is a piecewise exponential fit to the data, with a knot at $\Omega_{\text c, ex} = 0.20(3)\Er$.
The vertical dotted line indicates the theoretically predicted $\Omega_{\text c, th}=0.21\Er$, while the shaded area indicates uncertainty around $\Omega_{\text c, ex}$ from the piecewise exponential fit.
}
\label{fig:sm:Ome_Thold_phDiag}
\end{figure}

We continue by investigating the time-scale for equilibrium at $\delta=0$ by varying the hold time $\th$. 
As $\th$ increased, the BEC either remained phase-mixed (in the stripe phase, $\Wh\lesssim\Wc$) or phase-separated (in the plane-wave phase, $\Wh\gtrsim\Wc$).
As shown in the inset of Fig.~\ref{fig:sm:Ome_Thold_phDiag}, the Bragg signal $N_{\rm B}\left(\th\right)$ decayed with different rates in the stripe phase versus the plane-wave phase, and the individual time-traces are well fit by exponentials (the solid curves).  
We attribute the slowly decreasing Bragg signal in the stripe phase to a combination of loss processes and redistribution of the in-trap spin distribution as governed by the $c_2$ intra-spin interaction between $\ket{\uparrow}$ and $\ket{\downarrow}$, while we associate the quickly dropping Bragg signal to the rapid formation of domain structure. 

Figure~\ref{fig:sm:Ome_Thold_phDiag} shows the resulting time constants $\tau$ as a function of $\Wh$.
We note that the time constant is typically in excess of $\unit[50]{ms}$ in the stripe phase, and nearly plateaus in the plane-wave phase. 
We quantify this lifetime in terms of a piecewise exponential model with a knot at  $\Omega_{\text c, ex}$.  
We associate transition between the stripe and plane-wave phases with the intersection of the two component exponentials at $\Omega_{\text c, ex}=0.20(3)\Er$, indicated by the shaded area in Fig.~\ref{fig:sm:Ome_Thold_phDiag} and in agreement with the expected transition strength $\Wc=0.21\Er$.

\section{Whole-system scattering model}

Here we describe our simple 1D model of Bragg scattering from a spin-orbit coupled BEC, both in the stripe phase (in which the whole cloud contributes to the Bragg signal) and in the plane-wave phase (in which case only a domain wall separating the spin components contributes).  In our experiments, we ramp the Raman coupling to $\approx 7 \Er$ just prior to all our Bragg pulses, removing any explicit dependance of the Raman coupling strength from the Bragg signal.  

For this simple model, we neglect the very small difference in the interaction strengths $g_{\uparrow,\uparrow}$ and $g_{\downarrow,\downarrow}$, and describe the interactions in terms of a large spin-independent term $c_0$ and a weak spin-dependent contribution $c_2$.  We focus on a 1D system with Thomas-Fermi radius $R$.

For $\Omega>\Wc$, the spin-healing length is
$$
\xi_s = \left(\frac{\hbar^2}{2 m c_2 n}\right)^{1/2} = \left(\frac{c_0}{c_2}\right)^{1/2}\left(\frac{\hbar^2}{2 m \mu}\right)^{1/2} = \left(\frac{c_0}{c_2}\right)^{1/2} \xi;
$$
where $n$ is the local density; $\mu=c_0 n$ is the chemical potential; and $\xi$ is the conventional healing length.

For small coupling strengths $\Omega$, we introduce a spin-dependent interaction
$$
c_2^\prime = c_2 + c_0\frac{\Omega_R^2}{8 E_R^2},
$$
and in the case of $^{87}{\text {Rb}}$, we have $-c_2 / c_0\approx 0.005$.  This expression can be simplified in terms of a critical coupling $\Wc^2 = (-c_2/c_0) 8 E_R^2$ giving
$$
c_2^\prime = (-c_2) \left(\frac{\Omega_R^2}{\Omega_C^2}-1\right),
$$
For equal spin populations and for $\Omega>\Wc$, a domain wall forms at position $x=0$ leading to the spin densities
$$
n_{\uparrow,\downarrow} =  n_0 \frac{1 \pm \tanh(x/\xi_s)}{2}.
$$
In a SOBEC, the system exhibits density modulations
$$
\delta n(x) \propto \sqrt{n_\uparrow(x) n_\downarrow(x)} = n_0 \frac{\text{sech}(x/\xi_s)}{2},
$$
and the scattered electric field is
\begin{align*}
E_{\text B} &\propto \int_{-R}^{R} d x \sqrt{n_\uparrow(x) n_\downarrow(x)}\\
 &= 2 n_0 \xi_s \tan^{-1} \left[\tanh \left(\frac{R}{2 \xi_s}\right)\right].
\end{align*}
This yields the scattered intensity 
$$
\frac{I_{\text B}}{I_{\rm max}} =\frac{1}{\alpha^2} \left\{\tan^{-1} \left[\tanh \left(\alpha\right)\right]\right\}^2,
$$
normalized to the fully mixed case in terms of the dimensionless parameter $\alpha = R / 2 \xi_s$.

Now, specializing to the SOBEC case, with $c_2\rightarrow c_2^\prime$ we obtain the final fitting function
\begin{align*}
\frac{I_{\text B}}{I_{\rm max}} =&  \frac{1}{\alpha^2}\left(\frac{\Omega_R^2}{\Omega_C^2}-1\right)^{-1} \times\\
& \left(\tan^{-1} \left\{\tanh \left[\alpha\left(\frac{\Omega_R^2}{\Omega_C^2}-1\right)^{1/2}\right]\right\}\right)^2,
\end{align*}
which approaches $1$ as $\Omega\rightarrow \Omega_C$, as expected at the transition from the plane-wave phase to the stripe phase.

The integral in Eq.~(S1) assumed a uniform density distribution, whereas the harmonic confinement in all three spatial directions gives the overall 1D density $n(x) = n_0 [1 - (x/R)^2]^{2}$.  In addition, the reduced average density in the transverse dimension increases the healing length by a factor of $\sqrt{2}$ when expressed in terms of the peak density.

\begin{figure}
\begin{centering}
\includegraphics{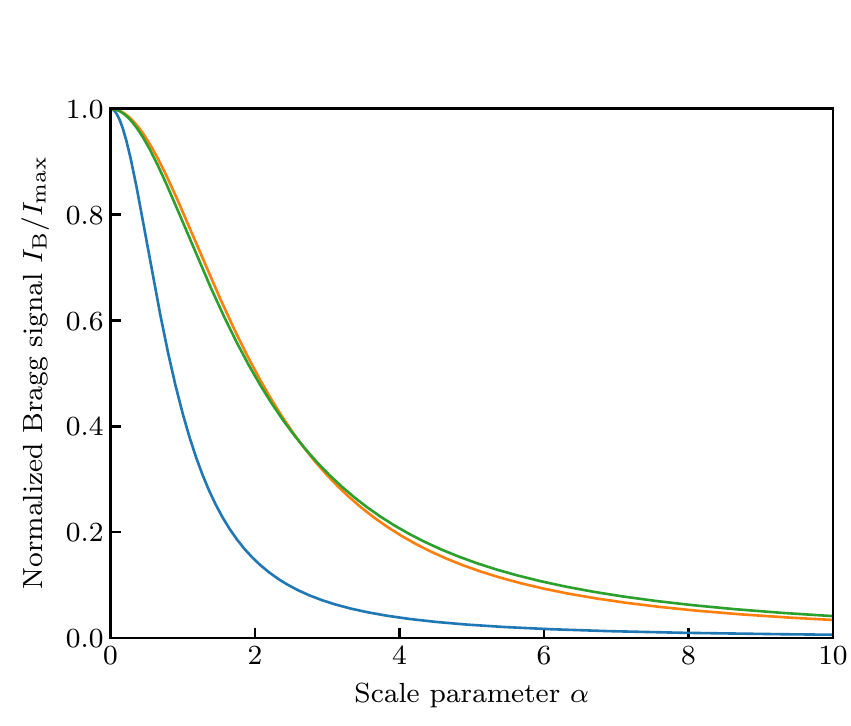}
\end{centering}
\caption{
Bragg scattering models.  Blue: uniform density model.  Orange: Thomas-Fermi model.  Green: uniform density model with $\alpha$ scaled by a factor of $2.35$.
} 
\label{fig:sm:models}
\end{figure}

We evaluated (S1) shaped by this profile, and the result is an unenlightening sum of trigonometric and polylog functions, still parameterized by $\alpha$ and $\Omega_R/\Omega_C$.  Figure~\ref{fig:sm:models} plots three different cases for these models: the uniform density model (blue), the Thomas-Fermi model (orange), and the uniform density model with $\alpha$ scaled by a factor of $2.35$ (green).  We see that the scaled uniform density model nearly overlaps the Thomas-Fermi model, and for ease of calculation we use this scaled model for our fitting function in the main document.

\bibliographystyle{apsrev4-1}
\bibliography{StripePhase}